\documentclass[12pt]{article}
\usepackage{graphicx}
\usepackage{amsmath}
\usepackage{amssymb}
\usepackage{authblk}

\newcommand{\mystackrel}[2]{%
  \setbox0 = \vbox{\hbox{$#2$}} %
  \stackrel{#1}{#2}\kern -3pt %
  \vbox to \ht0 {%
    \hbox to 1pt{} %
  } %
}
\newcommand{\nablaflat}{%
  \setbox0 = \vbox{\hbox{$\nabla$}} %
  \stackrel{\rm\tiny flat}{\nabla}\kern -5pt %
  \vbox to \ht0 {%
    \hbox to 1pt{} %
  } %
}

\def\amshow{0}

\ifodd\amshow

\usepackage{hyperref}

\fi

\title{A construction of integrated vertex operator in the 
pure spinor sigma-model in $AdS_5\times S^5$}

\author{
Osvaldo Chandia\thanks{Departamento de Ciencias, 
Falcultad de Artes Liberales,\\ Facultad de Ingenier\'ia y Ciencias,
Universidad Adolfo Ib\'a\~nez, Santiago, Chile
}\;,
Andrei Mikhailov\thanks{
Instituto de F\'{i}sica Te\'orica, Universidade Estadual Paulista\\
R. Dr. Bento Teobaldo Ferraz 271, S\~{a}o Paulo, Brasil
}\;
and
Brenno C. Vallilo\thanks{
 Departamento de Ciencias F\'isicas, Facultad de Ciencias Exactas, \\
Universidad Andres Bello, Republica 220, 
Santiago, Chile
}

}

\date{}

\begin{document} 

\maketitle

\begin{abstract}
Vertex operators in string theory come in two varieties: integrated and 
unintegrated. Understanding both types is important for the calculation of
the string theory amplitudes. The relation between them is a descent procedure
typically involving the $b$-ghost. In the pure spinor formalism vertex 
operators can be identified as cohomology classes of an infinite-dimensional
Lie superalgebra formed by covariant derivatives. We show that in this 
language the construction of the integrated vertex from an unintegrated
vertex is very straightforward, and amounts to the evaluation  of the cocycle
on the generalized Lax currents. 
\end{abstract}

\section{Introduction and notations}

\subsection{Introduction}
It is a crucial fact, that string worldsheet theories come in families. In 
other words, they are deformable. The infinitesimal deformations of the 
worldsheet action $S$ are parametrized by {\em integrated vertex operators} $U$:
\begin{equation}\label{DeformationOfTheAction}
   S\mapsto S + \varepsilon \int d\tau d\sigma\; U
\end{equation}
For every integrated vertex operator, there is the corresponding unintegrated
one, usually called $V$. The unintegrated vertices of the Type II theory carry 
ghost number two. They correspond to the cohomology of the BRST operator
$Q_{\rm BRST}$:
\begin{align}
   Q_{\rm BRST}V = \;& 0
\label{Closed}
\\    
V\simeq \;& V + Q_{\rm BRST}\Phi
\label{Exact}
\end{align}
In the pure spinor formalism, the unintegrated vertex operators are, 
schematically, of the form:
\begin{equation}
   V = A^{LR}_{\alpha\dot{\alpha}}(x,\theta)\lambda_L^{\alpha}\lambda_R^{\dot{\alpha}}
+ A^{LL}_{\alpha\beta}(x,\theta)\lambda_L^{\alpha}\lambda_L^{\beta}
+ A^{RR}_{\dot{\alpha}\dot{\beta}}(x,\theta)
\lambda_R^{\dot{\alpha}}\lambda_R^{\dot{\beta}}
\end{equation}
where $x,\theta$ are the spacetime coordinates and $\lambda_{L},\lambda_R$ the pure spinor ghosts.

Understanding both integrated and unintegrated versions of the vertex 
operators is important for the string perturbation theory \cite{Berkovits:2000fe}. The unintegrated
vertices are usually somewhat simpler. They correspond to the
cohomology classes of (\ref{Closed}), (\ref{Exact}). Their construction does not require the
use of the worldsheet equations of motion. 

The relation between $U$ and $V$ was explained in \cite{Berkovits:2000ph,Berkovits:2001ue}. Given the integrated 
vertex $V =A_{\alpha\dot{\alpha}}(x,\theta)\lambda_L^{\alpha}\lambda_R^{\dot{\alpha}}$, one can construct the integrated vertex as a 
linear combination of the expressions like 
$A_{\alpha\dot{\alpha}}\partial_+\theta_L^{\alpha}\partial_-\theta_R^{\dot{\alpha}}$, $A_{mn}\Pi^m_+\Pi^n_-$, $\Omega_{mn\;\hat{\alpha}}N^{mn}_+\partial_-\theta^{\hat{\alpha}}_R$, $S_{mnpq}N^{mn}_+N^{pq}_-$
{\it etc.}, where $A,\Omega,S,\ldots$ are superfields 
({\it i.e.} functions of $x$ and $\theta$) constructed from the $A_{\alpha\dot{\alpha}}$ by evaluating 
various derivatives. The precise relation is given by the 
``descent procedure'': 
\begin{align}
   dV = \;& QV^{(1)}
\label{DescentV1}\\    
\label{DescentU}dV^{(1)} =\;& QU
\end{align}
However, this descent procedure is somewhat mysterious\footnote{The descent 
procedure should be controlled by the $b$-ghost \cite{Aisaka:2009yp}. But the pure
spinor $b$-ghost is somewhat problematic, being a composite field with 
denominators (see \cite{Chandia:2010ix,Jusinskas:2013yca,Berkovits:2013pla} for the recent progress and references therein).} and to the 
best of our knowledge there was no formal proof that it always works. 

In this note we will give some mathematical interpretation of this descent
procedure. We will consider both flat spacetime ${\bf R}^{1+9}$ and $AdS_5\times S^5$, but 
mostly concentrate on $AdS_5\times S^5$ (because flat spacetime is a limit of AdS). 
The unintegrated vertices were interpreted in \cite{Mikhailov:2012uh} as elements of the relative
Lie algebra cohomology for some infinite-dimensional Lie superalgebra ${\cal L}_{\rm tot}$.
It turns out that in this language, the relation between $U$ and $V$ is rather
straightforward. The generalization of the Lax pair considered in \cite{Mikhailov:2013vja} plays 
the key role in the construction. Our construction proves the existence of
the chain of operators participating in the descent procedure ({\it i.e.},
given the unintegrated operator $V$, always exist $V^{(1)}$ and $U$ satisfying
(\ref{DescentV1}) and (\ref{DescentU})). It implies that the relation between the unintegrated and 
integrated vertex operators can be understood as a manifestation of the Koszul 
duality\footnote{
An excellent review of the Koszul duality in the context of pure spinor 
formalism can be found in the introductory sections of \cite{Gorodentsev:2006fa}.}.

\subsection{Plan of the paper}
We will start in Section \ref{sec:IntVertexClosed} by constructing the integrated vertex operator 
for the closed Type IIB superstring in $AdS_5\times S^5$ (our most general case). 
Then in Section \ref{sec:BetaDef} we give a simple example (the $\beta$-deformation) where 
everything can be done explicitly. The flat space limit is considered in 
Section \ref{sec:FlatSpace}. Finally, in Section \ref{sec:Maxwell} we will consider the construction of the 
integrated vertex for the open superstring.

\subsection{Notations}
Let ${\bf g} = {\bf psu}(2,2|4)$ denote the supersymmetry algebra of $AdS_5\times S^5$ and $U{\bf g}$ 
its universal enveloping algebra. The space of formal Taylor series of 
complex-valued functions at the unit ${\bf 1}\in PSU(2,2|4)$ can be identified with 
the dual space $(U{\bf g})'= \mbox{Hom}(U{\bf g},{\bf C})$.  We denote ${\bf g}_{\bar{0}} = so(1,4)\oplus so(5)$ the
subalgebra leaving invariant a fixed point in $AdS_5\times S^5$. Then the space
of Taylor series of functions in $AdS_5\times S^5$ is $\mbox{Hom}_{U{\bf g}_{\bar{0}}}(U{\bf g},{\bf C})$.

The infinite-dimensional Lie superalgebra ${\cal L}_{\rm tot}$ was introduced in \cite{Mikhailov:2012uh}. 
It contains an ideal $I\subset {\cal L}_{\rm tot}$ such that ${\cal L}_{\rm tot}/I = {\bf g}$. Let $\pi$ denote the
projection:
\begin{align}
   \pi\;:\; {\cal L}_{\rm tot} \to {\bf g}
\end{align}

\section{Integrated vertex for closed string}\label{sec:IntVertexClosed}
In this section we will review the interpretation of the unintegrated vertex 
in terms of the relative cohomology of ${\cal L}_{\rm tot}$, and then explain how to use this
language to describe the integrated vertex and the descent procedure.

\subsection{Unintegrated vertices from the relative cohomology}
Unintegrated vertices correspond to the elements of  $H^2({\cal L}_{\rm tot},{\bf g}_{\bar{0}}, \; (U{\bf g})')$ \cite{Mikhailov:2012uh}.
The 2-cocycle
representing such an element is a bilinear function of two elements of ${\cal L}_{\rm tot}$, 
which we will denote $\xi$ and $\eta$. We have to remember that the 2-cocycle takes
values in $(U{\bf g})'$ which is identified with the space of Taylor series at the
unit of the group manifold. Let $g\in PSU(2,2|4)$ denote the group element.
The value of the 2-cocycle is a Taylor series of $g$, not necessarily 
convergent\footnote{The ${\bf g}_{\bar{0}}$-covariance condition in the definition of the relative cohomology implies that this is actually
a section of a bundle over $AdS_5\times S^5$}. Therefore, the element $\psi\in H^2({\cal L}_{\rm tot},{\bf g}_{\bar{0}}, \; (U{\bf g})')$ is a function
of two ghosts $\xi$ and $\eta$ taking values in the functions of $g$:
\begin{equation}
   \psi(\xi,\eta)(g)
\end{equation}
To understand the cocycle condition on $\psi$, we introduce the left derivative.
For $a\in {\bf g}$ and $f\in (U{\bf g})'$:
\begin{align}
   (L(a) f)\;(g) = & \left.{d\over ds}\right|_{s=0} f(e^{sa}g)
\end{align}
We have $[\;-L(a)\;,\;-L(b)\;]\;=\; -L([a,b])$. 
With this definition, the cocycle condition on $\psi$, or vanishing of 
$(Q_{\rm Lie}\psi)(\zeta,\xi,\eta)$, is:
\begin{align}\label{CocycleCondition}
L(\pi(\zeta)) \psi(\xi,\eta) 
 + \psi([\zeta,\xi],\eta) \; + 
\\    
+ \;\mbox{cycl}(\zeta,\xi,\eta) = 0
\nonumber
\end{align}
where $\pi$ is the projection from ${\cal L}_{\rm tot}$ to ${\bf g}$.

\subsection{Ansatz for integrated vertex}
Let $\psi$ be the cocycle representing an element of $H^2({\cal L}_{\rm tot},{\bf g}_{\bar{0}}, \; (U{\bf g})')$. As
we have explained, it corresponds to a function of two ghosts $\xi,\eta$ and a group 
element $g$:
\begin{align}
   \psi(\xi,\eta)(g)
\end{align}
The Lax operator $L_{\pm}$ was introduced in \cite{Mikhailov:2012uh}.
Let us define $J_{\pm}\in {\cal L}_{\rm tot}$ starting from the Lax operator:
\begin{equation}
   L_{\pm} = {\partial\over\partial\tau^{\pm}} + J_{\pm} 
\end{equation}
The BRST transformation of $J_{\pm}$ was calculated in \cite{Mikhailov:2012uh}:
\begin{align}
   \epsilon Q_{\rm BRST}J_{\pm}= -[L_{\pm}\;,\;
   \epsilon\lambda_L^{\alpha}\nabla^L_{\alpha} + 
   \epsilon\lambda_R^{\dot{\alpha}}\nabla^R_{\dot{\alpha}} ]
\end{align}
We will denote:
\begin{align}
   \Lambda =    \lambda_L^{\alpha}\nabla^L_{\alpha} + 
   \lambda_R^{\dot{\alpha}}\nabla^R_{\dot{\alpha}} 
\end{align}
Let us consider the following ansatz for the integrated vertex:
\begin{align}
 U =  \psi(J_+,J_-)(g)
\end{align}
In the rest of this section we will prove that this $U$ satisfies the descent
equations (\ref{DescentU}), (\ref{DescentV1}). Therefore it is the integrated vertex, {\it i.e.} the
deformation of the action (\ref{DeformationOfTheAction}). It would be interesting to explicitly confirm
that the deformed action is conformally invariant in perturbation theory; we
leave this for the future work \cite{ChandiaMikhailovValliloToAppear}. 

\subsection{Ascent: the first step}
We have:
\begin{align}
   (\epsilon Q_{\rm BRST}\psi) (J_+,J_-) = &\;
- \psi\left(
   \left[
      {\partial\over\partial \tau^+} + J_+\;,\;\epsilon\Lambda
   \right]\;,\;J_-\right) \;-
\nonumber \\
&\; -
\psi\left(
   J_+\;,\;\left[
      {\partial\over\partial\tau^-} + J_-\;,\;\epsilon\Lambda
\right]\right) \;+
\nonumber \\   
&\; + L(\pi(\epsilon\Lambda)) \psi(J_+,J_-) \; = 
 \\[8pt]   
=&\;
- {\partial\over\partial \tau^+} 
\psi\left(\epsilon\Lambda\;,\;J_-\right)
+ {\partial\over\partial \tau^-}
\psi\left(\epsilon\Lambda\;,\;J_+\right) \;+
\nonumber \\
&\;
+ \psi\left(\epsilon\Lambda\;,\;\partial_+J_- - \partial_-J_+\right) \; -
\nonumber \\  
&\;
- L(\pi(J_{+}))\psi(\epsilon\Lambda\;,\;J_{-}) 
+ L(\pi(J_{-}))\psi(\epsilon\Lambda\;,\;J_{+}) \;+
\nonumber \\   
&\;
+ \psi([\epsilon\Lambda , J_+],J_-) 
+ \psi(J_+,[\epsilon\Lambda , J_-]) \;+
\nonumber \\  
&\; 
+ L(\pi(\epsilon\Lambda))\psi(J_+,J_-) \;
\end{align}
Taking into account that on-shell $\partial_+J_- - \partial_-J_+ = - [J_+,J_-]$ and $Q_{\rm Lie}\psi = 0$, 
we conclude that $(\epsilon Q_{\rm BRST}\psi) (J_+,J_-)$ on-shell is a total derivative:
\begin{align}
(\epsilon Q_{\rm BRST}\psi) (J_+,J_-) = \;&
- (Q_{\rm Lie}\psi)(\epsilon\Lambda,J_+,J_-) 
- \;\partial_+\psi(\epsilon\Lambda,J_-)\; 
+ \;\partial_-\psi(\epsilon\Lambda,J_+) =
\nonumber \\    
=\;&
- \;\partial_{+}\psi(\epsilon\Lambda,J_{-})\; 
+ \;\partial_-\psi(\epsilon\Lambda,J_+) 
\end{align}
\ifodd\amshow 
\includegraphics[scale=0.5]{snapshots/cocycle.png}

\fi
In other words:
\begin{equation}
\epsilon Q_{\rm BRST}\left(\psi(J_+,J_-) \;d\tau^+\wedge d\tau^- \right)\;=\;
- d\psi(\epsilon\Lambda,J)
\end{equation}

\subsection{Ascent: the second step}
At the first step we have seen that: 
\begin{equation}
\epsilon Q_{\rm BRST}U = -d\psi(\epsilon\Lambda,J)
\end{equation}
Continuing the ascent, we get:
\begin{align}
   \epsilon' Q_{\rm BRST}\left(\psi(\epsilon\Lambda,J)\right) =& - \psi\left(
      \epsilon\Lambda,
      \left[
         d + J\;,\;\epsilon'\Lambda
      \right]
   \right) + (L(\pi(\epsilon'\Lambda))\psi(\epsilon\Lambda,J)) =
\nonumber \\[8pt]   
=& - {1\over 2} d \left( 
   \psi(\epsilon\Lambda,\;\epsilon'\Lambda)
\right) - {1\over 2}(L(\pi(J))\psi(\epsilon\Lambda,\;\epsilon'\Lambda)) \;-
\nonumber \\ 
&\; - \psi\left(
      \epsilon\Lambda,
      \left[
         J\;,\;\epsilon'\Lambda
      \right]
   \right)\;+
\nonumber \\  
& \;+ (L(\pi(\epsilon'\Lambda))\psi(\epsilon\Lambda,J))\; =
\nonumber \\[8pt]    
=& - {1\over 2} d \left( 
   \psi(\epsilon\Lambda,\;\epsilon'\Lambda)
\right)
\label{AscentToUnintegrated}
\end{align}
Here we used that:
\begin{equation}
   \psi(\{\Lambda,\Lambda\},J)(g) = 0
\end{equation}
--- this is because $\psi$ represents a {\em relative} (w.r.to ${\bf g}_{\bar{0}}$) cohomology, 
and $\{\Lambda,\Lambda\}\in {\bf g}_{\bar{0}}$.

\ifodd\amshow 
\includegraphics[scale=0.5]{snapshots/cocycle-1.png}
\fi

Eq. (\ref{AscentToUnintegrated}) agrees with $\psi(\Lambda,\Lambda)(g)$ being the unintegrated vertex operator
corresponding to $\psi \in H^2({\cal L}_{\rm tot},\;{\bf g}_{\bar{0}},\;(U{\bf g})')$.

\subsection{If $\psi$ is exact, then $U$ is a total derivative}
On the other hand, consider the case when $\psi=Q_{\rm Lie}\phi$. In this case:
\begin{equation}
   \psi(\xi,\eta) = - L(\pi(\xi))\phi(\eta) + L(\pi(\eta))\phi(\xi) 
- \phi([\xi,\eta])
\end{equation}
And the corresponding integrated vertex is:
\begin{align}
   U = \;& - L(\pi(J_+))\phi(J_-) + L(\pi(J_-))\phi(J_+) - \phi([J_+,J_-])\; =
\nonumber  \\  
=\;& - L(\pi(J_+))\phi(J_-) + L(\pi(J_-))\phi(J_+) + 
\phi(\partial_+J_- - \partial_-J_+) \; =
\nonumber \\   
=\;& \partial_+\phi(J_-) - \partial_- \phi(J_+)
\end{align}

\section{Example: $\beta$-deformation}\label{sec:BetaDef}
The unintegrated vertex is: 
\begin{equation}\label{BetaUnintegrated}
V(\epsilon,\epsilon') =   
\left(g^{-1}(\epsilon\lambda_3-\epsilon\lambda_1)g\right)_a
B^{ab} 
\left(g^{-1}(\epsilon'\lambda_3-\epsilon'\lambda_1)g\right)_b
\end{equation}
where $B^{ab}$ is a constant antisymmetric tensor: $B\in ({\bf g}\wedge {\bf g})'$ \cite{Mikhailov:2009rx,Bedoya:2010qz}.

The corresponding element of $H^2({\cal L}_{\rm tot},{\bf g}_{\bar{0}},(U{\bf g})')$ is:
\begin{align}\label{PsiForBeta}
   \psi(\xi,\eta)(g) = 
   \left(
      g^{-1}\pi([\mbox{deg}\;,\;\xi]) g
   \right)_a
   B^{ab}
      \left(
      g^{-1}\pi([\mbox{deg}\;,\;\eta]) g
   \right)_b
\end{align}
Here $\mbox{deg}$ is the degree operator, which is an outer derivation of ${\cal L}_{\rm tot}$:
\begin{align}
   [\mbox{deg}\;,\;t^0_{[mn]}] =\;& 0
\nonumber\\  
[\mbox{deg}\;,\;\nabla^L_{\alpha}] = \;& \nabla^L_{\alpha}
\nonumber\\  
[\mbox{deg}\;,\;\nabla^R_{\dot{\alpha}}] = \;& - \nabla^R_{\dot{\alpha}}
\end{align}
In particular, 
\begin{equation}
   \left[
      \mbox{deg}\;,\;
      \lambda_3^{\alpha}\nabla^L_{\alpha}  + 
      \lambda_1^{\dot{\alpha}}\nabla^R_{\dot{\alpha}}
   \right] = 
      \lambda_3^{\alpha}\nabla^L_{\alpha}  -
      \lambda_1^{\dot{\alpha}}\nabla^R_{\dot{\alpha}}
\end{equation}
--- this ``explains'' the minus sign in (\ref{BetaUnintegrated}). It is straightforward
to verify that $\psi(\xi,\eta)(g)$ defined by (\ref{PsiForBeta}) satisfies the cocycle condition
(\ref{CocycleCondition}). In fact, the relative cohomology has a multiplicative structure, and
our $\psi$ is the product of 1-cocycles of the form $\xi\mapsto g^{-1}[\mbox{deg}\;,\;\xi]g$. 

In order to obtain the integrated vertex, we substitute $\xi\mapsto J_+$ and $\eta\mapsto J_-$:
\begin{align}
   U = \left(
      g^{-1}\pi([\mbox{deg}\;,\;J_+])g
   \right)_a B^{ab}\left(
      g^{-1}\pi([\mbox{deg}\;,\;J_-])g
   \right)_b
\label{IntegratedBeta}
\end{align}
This is equivalent to the expression $j_{a+}B^{ab}j_{b+}$ of \cite{Bedoya:2010qz}.

\section{Flat space limit}\label{sec:FlatSpace}
\subsection{The \.In\"on\"u-Wigner contraction}
The flat space limit of ${\cal L}_{\rm tot}$ is by the \.In\"on\"u-Wigner contraction, 
which is essentially a change of variables: 
\begin{align}
& \nablaflat^L_{\alpha} = \varepsilon\nabla^L_{\alpha} \;\;,\;
\nablaflat^L_m = \varepsilon^2\nabla^L_m\;,\;\;
\mystackrel{\rm flat}{W}^L_{\dot{\alpha}} = \varepsilon^3W^L_{\dot{\alpha}}\;,
\;\;\ldots
\label{GeneratorRescalingL} \\   
& \nablaflat^R_{\dot{\alpha}} = \varepsilon\nabla^R_{\dot{\alpha}}  \;\;,\;
\nablaflat^R_m = \varepsilon^2\nabla^R_m\;\;,\;
\mystackrel{\rm flat}{W}^R_{\alpha} = \varepsilon^3W^R_{\alpha}\;,
\;\;\ldots
\label{GeneratorRescalingR} \\   
\mbox{but: }& \mystackrel{\rm \tiny flat}{t}^0_{[mn]} =\; t^0_{[mn]}
\label{DoNotRescaleT0}
\end{align}
where $\varepsilon \to 0$ is a small parameter. In the limit $\varepsilon \to 0$: $\{\nablaflat^L_{\alpha}\;,\;\nablaflat^R_{\dot{\alpha}}\} = 0$.
Therefore, in this limit ${\cal L}_{\rm tot}$ becomes a semidirect sum:
\begin{equation}\label{BecomesDirectSum}
   ({\cal L}_L \oplus {\cal L}_R) \;+\; {\bf g}_{\bar{0}}
\end{equation}
where the ideal is the direct sum ${\cal L}_L \oplus {\cal L}_R$. The cohomology group 
$H({\cal L}_{\rm tot},{\bf g}_{\bar{0}}, (U{\bf g})')$ becomes $H({\cal L}_L \oplus {\cal L}_R\;,\; (U({\bf g}^{\rm \tiny flat}/{\bf g}_{\bar{0}}))')$. Notice that ${\bf g}^{\rm\tiny flat}/{\bf g}_{\bar{0}}$ 
is the flat space supersymmetry algebra $\bf susy$. We conclude that the flat space
vertex operators are identified with $H^2({\cal L}_L\oplus {\cal L}_R\;,\;(U({\bf susy}))')$.

Notice that $U({\cal L}_L\oplus {\cal L}_R)$ is a quadratic algebra, the Koszul dual to the
algebra of functions of two pure spinors $\lambda_L$ and $\lambda_R$. In this case the 
identification of the BRST cohomology with $H^2({\cal L}_L\oplus {\cal L}_R\;,\;(U({\bf susy})'))$ is a 
straightforward consequence of the central fact in the theory of Koszul
quadratic algebras, namely that the following complex:
\begin{align}
\ldots & \longrightarrow
   \mbox{Hom}_{\bf C}(A_n^!,A) 
\longrightarrow    
\mbox{Hom}_{\bf C}(A_{n-1}^!,A)
\longrightarrow\ldots
\nonumber\\     
\ldots & \longrightarrow
   \mbox{Hom}_{\bf C}(A_1^!,A) 
\longrightarrow A \longrightarrow {\bf C} \longrightarrow 0
\end{align}
provides a free resolution\footnote{we are greatful to the JHEP referee for 
pointing out an error in the original verision of this paragraph} of the trivial $A$-module $\bf C$. This means, that for
any module $V$ the cohomology $H^n(A, \;V)$ can be computed as $\mbox{Ext}^n_A({\bf C}, V)$, 
{\it i.e.} as the cohomology of the complex:
\begin{align}
   \ldots & \longrightarrow \mbox{Hom}_A(A,V) \longrightarrow 
   A_1^!\otimes\mbox{Hom}_A(A, V) \longrightarrow \ldots
\nonumber\\    
\ldots & \longrightarrow A_{n-1}^!\otimes \mbox{Hom}_A(A,V)
\longrightarrow A_{n}^!\otimes \mbox{Hom}_A(A,V)\longrightarrow \ldots
\end{align}
In our case  $A^!$ is the algebra of functions of $\lambda_L,\lambda_R$, $A = U({\cal L}_L\oplus {\cal L}_R)$, 
$V= (U({\bf susy}))'$. Also notice that $\mbox{Hom}_A(A,V) \simeq V$. The case of $AdS_5\times S^5$ 
is more subtle, because ${\cal L}_{\rm tot}$ is not a quadratic algebra.

\subsection{Lax pair in flat space}
Flat space can be described by the coset sPoincar\'e/Lorentz. An element of 
the AdS coset is $g=e^{X^m t^2_m + \theta^\alpha_L t^3_\alpha + \theta^{\dot\alpha} t^1_{\dot\alpha}}$. To go to the flat space limit we do the
field redefinition:
\begin{align}
&   \theta_L^{\alpha} = \varepsilon \vartheta_L^{\alpha}\;,\;\;
   \theta_R^{\hat{\alpha}} = \varepsilon \vartheta_R^{\hat{\alpha}}\;,\;\;
   \lambda_L^{\alpha} = \varepsilon \tilde{\lambda}_L^{\alpha}\;,\;\;
   \lambda_R^{\dot{\alpha}} = \varepsilon \tilde{\lambda}_R^{\dot{\alpha}}
\nonumber \\    
&   X^m = \; \varepsilon^2 x^m 
\nonumber \\   
&  {\partial L\over \partial (\partial_+\theta_L)} = 
\varepsilon^3 p_{L\alpha -} \;,\;\;
   {\partial L\over \partial (\partial_-\theta_R)} = 
\varepsilon^3 p_{R\alpha +} \;,\;\;
w_{1+} = \varepsilon^3 \tilde{w}_{1+}\;,\;\;
w_{3-} = \varepsilon^3 \tilde{w}_{3-}
\end{align}
Let us study the behaviour of the Maurer-Cartan currents under such a 
rescaling. They can be decomposed according to the ${\bf Z}_4$-grading
\begin{align}
- K_+=\;& \partial_+ g \;g^{-1}= 
\nonumber \\   
=\;&  \varepsilon \left(
   \partial_+ \vartheta^\alpha_L + O(\varepsilon^2)
\right)t^3_\alpha 
+ \varepsilon^2\left(
   \Pi^m_+ + O(\varepsilon^2)
\right)t^2_m + \varepsilon^3 d_{+}^{\dot{\alpha}}t^1_{\dot\alpha}\;
- \varepsilon^4 \mystackrel{\rm flat}{K}_{\bar{0}+}
\label{KPlus}\\[8pt]    
- K_-=\;& \partial_- g \;g^{-1}= 
\nonumber \\  
=\;&
\varepsilon\left(
   \partial_-\theta^{\dot\alpha}_R + O(\varepsilon^2)
\right)t^1_{\dot\alpha}  + 
\varepsilon^2\left(
   \Pi^m_- + O(\varepsilon^2)
\right)t^2_m + 
\varepsilon^3d_-^{\dot{\alpha}}t^3_{\dot{\alpha}}
- \varepsilon^4 \mystackrel{\rm flat}{K}_{\bar{0}+}
\label{KMinus}
\end{align}
where\footnote{we are using the convention where there structure constants 
of the $\bf susy$ algebra do not have $1/2$ and $i$.}
\begin{align}
\Pi_+^m=\;& 
\partial_+ x^m + (\vartheta_L\gamma^m\partial_+\vartheta_L) 
\label{PiPlus}\\  
\Pi_-^m=\;& 
\partial_- x^m + (\vartheta_R\gamma^m\partial_-\vartheta_R)
\label{PiMinus}\\  
\varepsilon^3 d^{\dot{\alpha}}_+ = J_{1+}^{\dot{\alpha}}=\;& 
C^{\dot{\alpha}\alpha}d_{\alpha +}
\label{RaiseIndJPlus}\\   
\varepsilon^3 d^{\alpha}_- = J_{3-}^{\alpha}=\;& 
C^{\alpha\dot{\alpha}}d_{\dot{\alpha}-}
\end{align}
These $K_{\pm}$ satisfy the Maurer-Cartan identities:
$$\partial_+ K_- - \partial_-K_+ - [K_+,K_-]=0.$$
To construct $J_{\pm}$, we start with $K_{\pm}$ and do three things:
\begin{enumerate}
\item In the $K_+$ of (\ref{KPlus}), replace the ${\bf psu}(2,2|4)$ generators $t_{\alpha}^3$, $t_m^2$, $t_{\dot{\alpha}}^1$ with 
   $\nabla^L_{\alpha}$, $\nabla^L_m$ and $W^L_{\dot{\alpha}}$. Similarly, in the $K_-$ of (\ref{KMinus}) replace $t^1_{\dot{\alpha}}$, $t^2_m$ and $t^3_{\alpha}$ 
   with $\nabla_{\dot{\alpha}}^R$, $\nabla^R_m$ and $W_{\alpha}^R$.
\item Add  
   $\{\lambda_L^{\alpha} \nabla^L_{\alpha}\;,\;w_{L+}^{\dot{\beta}}W^L_{\dot{\beta}}\} - \lambda_L^{\alpha} w_{L+}^{\dot{\beta}} f_{\alpha\dot{\beta}}{}^{[mn]}t^0_{[mn]}$ to
   the $K_+$ and similarly add  
   $\{\lambda_R^{\dot{\alpha}} \nabla^R_{\dot{\alpha}}\;,\;w_{R-}^{\beta}W^R_{\beta}\} - \lambda_R^{\dot{\alpha}} w_{R-}^{\beta} f_{\dot{\alpha}\beta}{}^{[mn]}t^0_{[mn]}$ to the $K_-$.
\item Finally, replace the generators of ${\cal L}_{\rm tot}$ with the rescaled generators using 
   (\ref{GeneratorRescalingL}), (\ref{GeneratorRescalingR}) and (\ref{DoNotRescaleT0}) and drop all the terms containing positive powers 
   of $\varepsilon$. 
\end{enumerate}
This results in the following flat space expressions:
\begin{align}
   J_+ =\;&
   \partial_+ \vartheta^\alpha_L \nablaflat^L_{\alpha}
   + \Pi^m_+\nablaflat^L_m 
   + d_{\alpha +}\mystackrel{\rm flat}{W}_L^{\alpha}
   + \lambda_L^{\alpha} w_{\beta +} 
   \{\nablaflat_{\alpha}^L\;,\;\mystackrel{\rm flat}{W}_L^{\beta}\}
\label{FlatJPlus}
   \\     
   J_- =\;&
   \partial_- \vartheta^{\dot{\alpha}}_R \nablaflat^R_{\dot{\alpha}}
   + \Pi^m_-\nablaflat^R_m 
   + d_{\dot{\alpha}-}\mystackrel{\rm flat}{W}_R^{\dot{\alpha}}
   + \lambda_R^{\dot{\alpha}} w_{\dot{\beta}-}
   \{\nablaflat_{\dot{\alpha}}^R\;,\;\mystackrel{\rm flat}{W}_R^{\dot{\beta}}\}
\label{FlatJMinus}
\end{align}
Notice that in the flat space limit the left generators $\nabla^L_{\alpha},\nabla^L_m,\ldots$ commute 
with the right generators $\nabla^R_{\dot{\alpha}}, \nabla^R_m,\ldots$ (see Eq. (\ref{BecomesDirectSum})). Therefore the zero 
curvature conditions become $\partial_+J_- - \partial_-J_+ = 0$ which is equivalent to
\begin{equation}
   \partial_+J_- = \partial_-J_+ = 0
\end{equation}
This is simply the statement that $\partial_+\vartheta_L, \Pi_+, d_+, \lambda_L, w_{L+}$ are holomorphic and 
$\partial_-\vartheta_R,\Pi_-,d_-,\lambda_R,w_{R-}$ are antiholomorphic. 

Using the flatspace BRST transformations it can also be shown that the 
generalized Lax pair has the following BRST transformation:
$$\epsilon Q^{\rm{flat}}_{BRST} J_{\pm}= [\epsilon\Lambda , L_{\pm}] = [\epsilon\Lambda ,\partial_{\pm} + J_{\pm}],$$
where $\Lambda= \lambda^\alpha_L \nabla^L_\alpha + \lambda^{\dot\alpha}_R \nabla^R_{\dot\alpha}$

\section{Construction of integrated vertex for the open string}\label{sec:Maxwell}
In the case of Maxwell theory, the 
Koszul duality implies that unintegrated vertices can be identified as elements
of $H^1({\cal L}_{\rm YM}, (U{\bf susy}_{N=1})')$, see \cite{Mikhailov:2012uh}. The $(U{\bf susy}_{N=1})'$ is identified with the 
space of Taylor series at the unit of the group manifold (the super Minkowski 
space). Therefore an element of $H^1({\cal
L}_{\rm YM}, (U{\bf susy}_{N=1})')$ is represented by a 
linear function of $\xi\in {\cal L}_{\rm YM}$ taking values in the Taylor series of 
$g = (x,\theta)\in {\rm
SUSY}_{N=1}$. As in Section \ref{sec:IntVertexClosed}, we will simply write:
\begin{equation}\label{OpenStringUnintegrated}
   \psi(\xi)(g)
\end{equation}
Consider the Maxwell theory living on a $D$-brane. Consider the flat space Lax 
current $J_+d\tau^+ + J_-d\tau^-$ from (\ref{FlatJPlus}), (\ref{FlatJMinus}), restricted on the string boundary.
On the boundary there is some relation of the form:
\begin{equation}
   d\theta_L^{\alpha} = A^{\alpha}_{\dot{\beta}} d\theta_R^{\dot{\beta}}
\end{equation}
and a similar relation for $x$. To construct the open string integrated vertex 
we substitute in (\ref{OpenStringUnintegrated}) in place of $\xi$ the restructed current $J_{\pm}$ with the 
replacement:
\begin{equation}
   \nablaflat^L_{\alpha} \mapsto \nabla_{\alpha}\;,\;
   \nablaflat^R_{\dot{\alpha}}\mapsto A_{\dot{\alpha}}^{\beta}\nabla_{\beta}\;,\ldots
\end{equation}
where $\nabla_{\alpha}$ are the generators of the super-Yang-Mills algebra.
We get the integrated boundary vertex operator:
\begin{align}
   U = \psi(J_+)(g)d\tau^+ + \psi(J_-)(g)d\tau^-
\end{align}
The cocycle condition on $\psi$ is:
\begin{equation}\label{CocycleXiEta}
   L(\pi(\xi))\psi(\eta) - L(\pi(\eta))\psi(\xi) + \psi([\xi,\eta]) = 0
\end{equation}
For example:
\begin{align}\label{CocycleDefAm}
   L(\pi(\nabla_{\alpha})) \psi(\nabla_{\beta}) + L(\pi(\nabla_{\beta})) \psi(\nabla_{\alpha})
+ \Gamma_{\alpha\beta}^m \psi(\nabla_m) = 0
\end{align}
In the literature on pure spinors $L(\pi(\nabla_{\alpha}))$ is usually denoted $D_{\alpha}$. Eq. (\ref{CocycleDefAm})
tells us to identify $\psi(\nabla_m)$ with $A_m$. We can continue writing the cocycle
conditions (\ref{CocycleXiEta}) for other values of $\xi$ and $\eta$. 
We have\footnote{we use the same letter $W$ for the generator of the 
super-Yang-Mills algebra and for the target space superfield}:
\begin{align}
\psi(\nabla_{\alpha})(x,\theta) =\;& A_{\alpha}(x,\theta)
\\    
\psi(\nabla_m)(x,\theta) = \;& A_m(x,\theta)
\\    
\psi(W^{\alpha})(x,\theta) = \;& W^{\alpha}(x,\theta)
\\    
\psi(\{\nabla_{\alpha},W^{\beta}\})(x,\theta) = \;& 
(\Gamma^{[mn]}F_{[mn]}(x,\theta))_{\alpha}^{\beta}
\\   
\ldots
\end{align}
This is the well-known chain of identities leading to the construction of the
integrated vertex:
\begin{equation}
U = A_{\alpha}(x,\theta)d\theta^{\alpha} + 
A_m(x,\theta)(dx^m + (\theta\Gamma^m d\theta)) +
W^{\alpha}d_{\alpha} + \left(\lambda\;\Gamma^{[mn]}F_{[mn]} w\right)
\end{equation}
The new thing here is that we give an interpretation of this chain in terms of 
the Lie algebra cohomology, and this actually explains why this chain exists 
and is self-consistent. 

\section*{Acknowledgments}
We would like to thank Nathan~Berkovits for discussions, and the JHEP referee 
for preparing a thorough review and pointing out a mistake in the original 
version. We thank the FONDECYT grant 1120263 for partial financial support.


\def\cprime{$'$} \def\cprime{$'$}
\providecommand{\href}[2]{#2}\begingroup\raggedright\endgroup

\end{document}